\documentclass{INTERSPEECH2024}

\usepackage{multirow}

\makeatletter
\newcommand*\concat
  {\mathbin{\mathmakebox[\widthof{${+}\m@th$}]{+\hskip-1emplus1fil+}}}

\makeatother

 \newcommand\citep{\cite}

\interspeechcameraready

\title{{XTTS: a Massively Multilingual Zero-Shot Text-to-Speech Model}
\vspace{-0.08cm}
}

\name{

Edresson Casanova$^{1*}$, Kelly Davis$^{2}$, Eren Gölge$^{3*}$, Görkem Göknar$^{2}$, Iulian Gulea$^{2}$, Logan Hart$^{3*}$, Aya Aljafari$^{1*}$, Joshua Meyer$^{2}$, Reuben Morais$^{4*}$,  Samuel Olayemi$^{2}$, and Julian Weber$^{3*}$}

\address{
    $^1$ Nvidia, $^2$ Coqui.ai, $^3$ Cantina.ai, $^4$ voize GmbH
    \thanks{* Most of the work was done at Coqui.ai.}
  }
\email{ecasanova@nvidia.com \vspace{-0.7cm}}


\begin{document}

\maketitle

\begin{abstract}
    {
     Most Zero-shot Multi-speaker TTS (ZS-TTS) systems support only a single language. Although models like YourTTS, VALL-E X, Mega-TTS 2, and Voicebox explored  Multilingual ZS-TTS they are limited to just a few high/medium resource languages, limiting the applications of these models in most of the low/medium resource languages. In this paper, we aim to alleviate this issue by proposing and making publicly available the XTTS system. Our method builds upon the Tortoise model and adds several novel modifications to enable multilingual training, improve voice cloning, and enable faster training and inference. XTTS was trained in 16 languages and achieved state-of-the-art (SOTA) results in most of them.
    }
\end{abstract}

\vspace{-0.12cm}
\section{Introduction}
\label{sec:intro}
Text-to-Speech (TTS) systems have received a lot of attention in recent years due to the great advances in deep learning. Most TTS systems were tailored from a single speaker's voice, but there is current interest in synthesizing voices for new speakers (not seen during training) employing only a few seconds of speech. This approach is called zero-shot multi-speaker TTS (ZS-TTS) as in \citep{jia2018transfer, choi2020attentron, casanova2021sc, yourtts, wang2023neural, jiang2023mega}.


 Monolingual ZS-TTS was first proposed by \cite{arik2018neural} which extended the DeepVoice~3 model~\cite{deepvoice3}. Meanwhile, Tacotron 2~\cite{tacotron2} was adapted using external speaker embeddings, allowing for speech generation that resembles the target speaker~\cite{jia2018transfer, cooper2020zero}. 
 SC-GlowTTS~\cite{casanova2021sc}  explored a flow-based architecture and improved voice similarity for unseen speakers in training with respect to previous studies while maintaining comparable quality. VALL-E \cite{wang2023neural} was the pioneer in exploring the language modeling approach for ZS-TTS. It is a text-conditioned language model trained on Encodec \cite{defossez2022high} tokens. Encodec encodes each audio frame with 8 codebooks at a 75Hz frame rate.  VALL-E improved voice similarity and naturalness for unseen speakers. Tortoise \cite{tortoise} also explored the language modeling approach for ZS-TTS. 
 It was trained with 49k hours of English speech and it achieved promising ZS-TTS performance, enhancing naturalness. StyleTTS 2 \cite{li2023styletts} was built upon the StyleTTS framework 
 and it leverages style diffusion and adversarial training with large speech-language models (e.g. WavLM \cite{chen2022wavlm}) to achieve human-level TTS and SOTA ZS-TTS performance.
 P-Flow \cite{kim2023p}  combines a prompted text encoder with a low-matching generative decoder to sample high-quality mel-spectrograms efficiently. P-Flow matches the speaker similarity performance of the VALL-E model with two orders of magnitude less training data and has more than $20\times$ faster sampling speed. HierSpeech++ \cite{lee2023hierspeech++} is an efficient hierarchical speech synthesis framework that consists of a hierarchical speech synthesizer, text-to-vec, and speech super-resolution model. To improve speaker similarity the authors introduced a bidirectional normalizing flow Transformer network using AdaLN-Zero. 
 To improve audio quality, they have proposed a dual-audio acoustic encoder to enhance the acoustic posterior. 
 HierSpeech++ achieved  ZS-TTS SOTA results, enhancing especially the generated audio quality.

Most ZS-TTS models support only a single language. However, there is current interest in training models in multiple languages, reducing the number of speech hours and speakers needed to have a ZS-TTS model in a target language. YourTTS \cite{yourtts} was the first multilingual ZS-TTS model. The authors proposed several changes to VITS model \cite{kim2021conditional} architecture to support multilingual training and ZS-TTS. The authors trained the model using approximately 1k speakers in the English language, 5 speakers in French, and 1 speaker in Portuguese. The model achieved SOTA results in the English language and promising results in the French and Portuguese languages. It can also do cross-lingual TTS producing a native accent in the target language. YourTTS model has shown the viability of training ZS-TTS models in scenarios where only a few speakers are available, enabling synthetic data generation for low-resource scenarios \cite{casanova23_interspeech}. VALL-E X \cite{zhang2023speak} was built upon VALL-E; however, the authors introduced a language ID to support multilingual TTS and speech-to-speech translation. VALL-E X can also do cross-lingual TTS, producing a native accent in the target language.  Mega-TTS 2 \cite{jiang2023mega} is a ZS-TTS model capable of handling arbitrary-length speech prompts. The model was trained on 38k hours of multi-domain language-balanced speech in English and Chinese.  Mega-TTS 2 achieved  SOTA performance with short speech prompts and also produced better results with longer speech prompts.
In parallel with our work, Voicebox \cite{le2023voicebox} was proposed. Voicebox is a non-autoregressive continuous normalizing flow model. In contrast to auto-regressive models (e.g. VALL-E), Voicebox can consume context not only in the past but also in the future. The Voicebox model was trained in 6 languages and it achieved SOTA results in cross-lingual ZS-TTS.

Although some papers explored multilingual ZS-TTS as in \cite{yourtts, zhang2023speak, le2023voicebox, jiang2023mega} the number of supported languages is still low. YourTTS model was trained with only three languages, VALL-E X and Mega-TTS 2 explored only two languages, and Voicebox explored six languages. Given that, the current ZS-TTS models are limited to a few medium/high resource languages, limiting the applications of these models in most of the low/medium resource languages. In this paper, we aim to solve this issue by proposing a massive multilingual ZS-TTS model that supports 16 languages, including English (en), Spanish (es), French (fr), German (de), Italian (it), Portuguese (pt), Polish (pl), Turkish (tr), Russian (ru), Dutch (nl), Czech (cs), Arabic (ar), Chinese (zh), Hungarian (hu), Korean (ko), and Japanese (ja).

The contributions of this work are as follows:
\begin{itemize}
 \vspace{-0.1cm}
    \item We introduced XTTS, a new multilingual ZS-TTS model that achieves SOTA results in 16 languages;
    \item XTTS is the first massively multilingual ZS-TTS model supporting low/medium resource languages;
    \item Our model can perform cross-language ZS-TTS without needing a parallel training dataset.
     \item XTTS model and checkpoints are publicly available at Coqui TTS\footnote{https://github.com/coqui-ai/TTS} and also on Hugging Face XTTS\footnote{https://huggingface.co/coqui/XTTS-v2/tree/v2.0.2} repository.
\end{itemize}


The audio samples for each of our experiments are available on the demo website\footnote{https://edresson.github.io/XTTS/}.

\vspace{-0.12cm}
\section{XTTS model}\label{sec:XTTS}

\begin{figure}[]
\centering
\resizebox{0.30\textwidth}{!}{%
\includegraphics[width=1\textwidth]{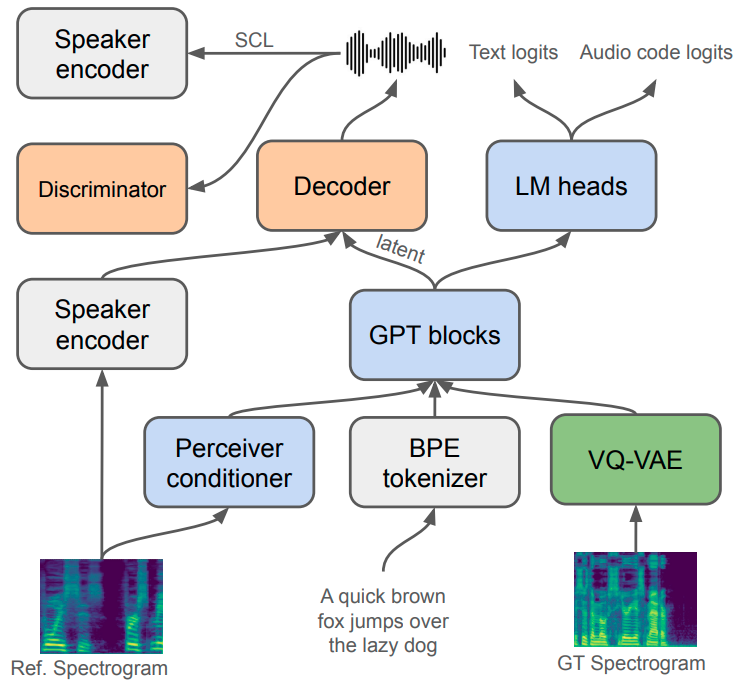}
}
 \vspace{-0.3cm}
 \caption{XTTS training architecture overview.}
  \vspace{-0.3cm}
 \label{fig:XTTS}
\end{figure}

XTTS builds upon Tortoise \cite{tortoise}, but includes several novel modifications to enable multilingual training, improve ZS-TTS, and enable faster training and inference. Figure \ref{fig:XTTS} shows an overview of the XTTS architecture. XTTS is composed of three components:

\textbf{VQ-VAE:} 
A Vector Quantised-Variational AutoEncoder (VQ-VAE) with 13M parameters receives a mel-spectrogram as input and encodes each frame with 1 codebook consisting of 8192 codes at a 21.53 Hz frame rate. The architecture and training procedure of VQ-VAE is the same as the one used in \cite{tortoise}; however, after VQ-VAE training we have filtered the codebook keeping only the first 1024 most frequent codes. In preliminary experiments, we verified that filtering the less frequent codes improved the model's expressiveness.

\textbf{Encoder:}
The GPT-2 encoder is a decoder-only transformer that is composed of 443M parameters, similar to \cite{tortoise}. It receives as inputs text tokens obtained via a 6681-token custom Byte-Pair Encoding (BPE) \cite{gage1994new} tokenizer and as output predicts the VQ-VAE audio codes. The GPT-2 encoder is also conditioned by a Conditioning Encoder, described below, that receives mel-spectrograms as input and produces 32 1024-dim embeddings for each audio sample.  The Conditioning Encoder is composed of six 16-head Scaled Dot-Product Attention layers followed by a Perceiver Resampler \cite{alayrac2022flamingo} to produce a fixed number of embeddings independently of the input audio length. Note that in \cite{tortoise} the authors didn't use the Perceiver Resampler, instead, they used only a single 1024-dim embedding to condition the GPT-2 encoder. In our preliminary experiments, we noticed that in massive multilingual training, the use of a single embedding leads to a decrease in the model's speaker cloning capability. 
We also have romanized the texts before tokenization for the Korean, Japanese, and Chinese languages using hangul-romanize\footnote{https://pypi.org/project/hangul-romanize/},  Cutlet\footnote{https://github.com/polm/cutlet}, and Pypinyin\footnote{https://pypi.org/project/pypinyin/} packages respectively.

\textbf{Decoder:}
The decoder is based on the HiFi-GAN vocoder \cite{kong2020hifi} with 26M parameters. It receives the latent vectors out of the GPT-2 encoder. Due to the high compression rate of the VQ-VAE, reconstructing the audio directly from the VQ-VAE codes leads to pronunciation issues and artifacts. To avoid this issue, we follow \cite{tortoise} and we have used the GPT-2 encoder latent space as input to the decoder instead of VQ-VAE codes. Our proposed decoder is also conditioned with speaker embedding from the H/ASP model \cite{heo2020clova}. The speaker embedding was added in each upsampling layer via linear projection. Inspired by \cite{yourtts}, to improve the speaker similarity, we also added the Speaker Consistency Loss (SCL). 

To speed up inference we have trained the VQ-VAE and the encoder using 22.5 kHz audio signals. However, we train the decoder by upsampling the input vectors linearly to the correct length to produce 24khz audio.  



\vspace{-0.12cm}
\section{Experiments}\label{sec:exp}

\vspace{-0.08cm}
\subsection{XTTS dataset}\label{sec:datasets}

The XTTS dataset is composed of public and internal datasets. Most of our internal data is in English and only public data is used for many languages. Table \ref{tab:XTTS-dataset} presents the number of hours for each language in the XTTS dataset. For English, we have used 541.7 hours from LibriTTS-R \cite{koizumi23_interspeech} and 1812.7 hours from  LibriLight \cite{kahn2020libri}. The rest of the English data was from the internal dataset that was composed of mostly audiobook-like data. For other languages, most of the data are from the Common Voice \cite{ardila2020common} dataset.


\begin{table}[]
\caption{Number of hours for each language in XTTS dataset.}
\label{tab:XTTS-dataset}
\centering
\resizebox{0.32\textwidth}{!}{%
\begin{tabular}{l|clc}
\hline
\textbf{Language} & \multicolumn{1}{c|}{\textbf{Hours}} & \multicolumn{1}{l|}{\textbf{Language}} & \textbf{Hours} \\ \hline
English        & \multicolumn{1}{c|}{14,513.1} & \multicolumn{1}{l|}{Czech}     & 52.4  \\ \hline
German         & \multicolumn{1}{c|}{3,584.4}  & \multicolumn{1}{l|}{Korean}    & 539.1 \\ \hline
Spanish        & \multicolumn{1}{c|}{1,514.3}  & \multicolumn{1}{l|}{Hungarian} & 62.0  \\ \hline
French         & \multicolumn{1}{c|}{2,215.5}  & \multicolumn{1}{l|}{Japanese}  & 57.3  \\ \hline
Italian        & \multicolumn{1}{c|}{1,296.6}  & \multicolumn{1}{l|}{Turkish}   & 165.3 \\ \hline
Portuguese      & \multicolumn{1}{c|}{2,386.8}  & \multicolumn{1}{l|}{Arabic}    & 240.9 \\ \hline
Russian        & \multicolumn{1}{c|}{147.1}    & \multicolumn{1}{l|}{Chinese}   & 233.9 \\ \hline
Dutch          & \multicolumn{1}{c|}{74.1}     & \multicolumn{1}{l|}{Polish}    & 198.8 \\ \hline
\textbf{Total} & \multicolumn{3}{c}{27,281.6}                                          \\ \hline
\end{tabular}
}
\vspace{-0.6cm}
\end{table}

\vspace{-0.08cm}
\subsection{Experimental setup}\label{sec:train}


Previous works \cite{li2023styletts, wang23c_interspeech, wang2023neural, kim2023p} that explored monolingual ZS-TTS have compared their models with the YourTTS model using the multilingual checkpoint released by the authors. This comparison is not fair because the number of hours of speech and the number of speakers are really important during ZS-TTS model training. Although the YourTTS multilingual model has been trained with more than 1k speakers in English, the model was trained with only 5 speakers in French and 1 speaker in Portuguese. Considering that the YourTTS authors have used a language batch balancer it means that during the training 66\% of the batch will be composed of samples from only 6 speakers. This can lead to overfitting reducing the performance in the English language (For more details see Section \ref{subsec:en-eval}). 

In this paper we have trained YourTTS on both LibriTTS \cite{zen2019libritts} and XTTS datasets to avoid these issues. In this way, we can compare YourTTS trained on only LibriTTS with current English ZS-TTS SOTAs. We can also compare it with the original multilingual YourTTS checkpoint to exhibit the problem with the comparison done in previous works. We can also fairly compare YourTTS trained with the XTTS dataset in 16 languages with our proposal model. For both XTTS and YourTTS trained with the XTTS dataset, we have used a language batch balancer.



We carried out three training experiments:
\begin{itemize}
     \item \textbf{Experiment 1:} YourTTS model trained only on English using LibriTTS train-clean-460 subset (the same data used in \cite{li2023styletts}) with the bug on SCL fixed\footnote{https://github.com/Edresson/YourTTS\#erratum}. We trained the model for 405k steps;
     \item \textbf{Experiment 2:} YourTTS trained on 16 languages using the XTTS dataset with SCL fixed for 1.96M steps;
    \item \textbf{Experiment 3:} XTTS model trained with the XTTS dataset for approximately 2.5M steps.
\end{itemize}

\vspace{-0.08cm}
\subsection{Training setup}\label{sec:train}


For YourTTS training we have used the Coqui TTS repository\footnote{https://github.com/coqui-ai/TTS}. XTTS and YourTTS were trained using an NVIDIA A100 with 80 GB GPUs. YourTTS experiments were run on a single GPU. XTTS was trained on 4 GPUs.

For the YourTTS generator training and for the discrimination of vocoder HiFi-GAN we use the AdamW optimizer with betas $0.8$ and $0.99$, weight decay $0.01$, and an initial learning rate of $0.0002$ decaying exponentially by a gamma of $0.999875$. We have used batch size equal to $64$. To speed up YourTTS experiments we used transfer learning from the checkpoints made publicly available at \cite{Cmltts2023}.

For XTTS training, we used the AdamW optimizer with betas $0.9$ and $0.96$, weight decay $0.01$, and an initial learning rate of $5e-05$ with a batch size equal to $4$ with grad accumulation equal to $16$ steps for each GPU. Following \cite{tortoise}, we only applied weight decay for weights and we also decayed the learning rate using MultiStepLR by a gamma of $0.5$ using the milestones $5000$, $150000$, and $300000$.

\vspace{-0.12cm}
\section{Results and Discussion}\label{sec:results}

We compared our model with the SOTAs ZS-TTS models: StyleTTS 2, Tortoise, YourTTS, HierSpeech++, and Mega-TTS 2. We also compared our model with a YourTTS model trained on our dataset for multilingual ZS-TTS. To make our work more reproducible, the evaluation code and all the audio samples are available at the ZS-TTS-Evaluation\footnote{https://github.com/Edresson/ZS-TTS-Evaluation} repository. 

To compare the models we have used 240 sentences for each supported language from FLORES+ \cite{nllb-22}. The sentences were chosen randomly from the $devtest$ subset. We have chosen the FLORES+ dataset because it has parallel translations for all languages supported by our model. In this way, we can compare all the language results using the same vocabulary. To test the ZS-TTS capability we decided to use all 20 speakers (10M and 10F) from the clean subset of the DAPS dataset\footnote{https://zenodo.org/records/4660670}. For each speaker, we randomly selected one audio segment between 3 and 8 seconds to use as a reference during the test sentence generation. We have used these samples to evaluate all languages, that way for non-English languages the models are compared in a cross-lingual way.

For YourTTS inference we have used a length scale equal to $1.0$, a noise scale equal to $0.3$, and a duration predictor noise scale equal to $0.3$. For XTTS inference we have used a temperature equal to $0.75$, length penalty equal to $1.0$, repetition penalty equal to $10.0$, top k equal to $50$, and top p equal to $0.85$. For Tortoise inference, we used the open-source available checkpoint with the parameters $num\_autoregressive\_samples$ equal to $256$, $diffusion\_iterations$ equal to $200$, and for the rest of the parameters we have used the default values. For StyleTTS 2, we have used the open-source checkpoint\footnote{https://github.com/yl4579/StyleTTS2\#inference} trained on the LibriTTS train-clean-460 subset, and for inference we have used the default parameters. For HierSpeech++, we have used the original model released by the authors on GitHub\footnote{https://github.com/sh-lee-prml/HierSpeechpp}, and for inference, we have used the default parameters.  For Mega-TTS 2, we have used samples kindly provided by the authors.

For the objective evaluation, following \cite{lee2023hierspeech++} we have used the UTMOS model \cite{saeki2022utmos} to predict the Naturalness Mean Opinion Score (nMOS). In \cite{lee2023hierspeech++}, the authors have used the open-source version of UTMOS\footnote{https://github.com/tarepan/SpeechMOS}, and the presented results of human nMOS and UTMOS are almost aligned. Although this can not be considered an absolute evaluation metric, it can be used to easily compare models in quality terms. To compare the similarity between the synthesized voice and the original speaker, we compute the Speaker Encoder Cosine Similarity (SECS) \cite{casanova2021sc} using the SOTA ECAPA2 \cite{thienpondt2024ecapa2} speaker encoder. 
Following previous works \cite{wang2023neural, kim2023p, lee2023hierspeech++}, we evaluate pronunciation accuracy using an ASR model. For it, we have computed the Character Error Rate (CER) using the Whisper Large v3 \cite{radford2022whisper} model. 

For subjective evaluation, we have measured user preference scores by comparing XTTS with previous models.

\vspace{-0.08cm}
\subsection{English evaluation}\label{subsec:en-eval}

\begin{table}[]
\caption{CER, UTMOS, and SECS for all our experiments and related works in the English language.}
\label{tab:results-en}
\centering
\resizebox{0.45\textwidth}{!}{%
\begin{tabular}{l|c|c|c|c}
\hline
\textbf{Model}   & \textbf{Hours}                                                         & \textbf{CER($\downarrow$)} & \textbf{UTMOS($\uparrow$)} & \textbf{SECS($\uparrow$)}\\ \hline
Ground truth & -  & -  & 4.2775 $\pm$ 0.15   & 0.8952 \\ \hline
Tortoise \cite{tortoise} & \begin{tabular}[c]{@{}l@{}}49k\end{tabular}  & 1.0934  &  4.0883 $\pm$ 0.31 & 0.5492 \\ \hline
StyleTTS 2 \cite{li2023styletts} & 245 & 0.6789  &   4.4260 $\pm$ 0.07 & 0.4728\\ \hline
Mega-TTS 2 \cite{jiang2023mega} & \begin{tabular}[c]{@{}l@{}}38k (all) \\ 27.5k (en)\end{tabular} & 1.4269  &   4.184 $\pm$ 0.17    &  0.6428 \\ \hline
HierSpeech++ \cite{lee2023hierspeech++} & 2.7k & 0.7741  &  \textbf{4.457 $\pm$ 0.06}    & 0.6530\\ \hline

\begin{tabular}[l]{@{}l@{}}Original  \\ YourTTS \cite{yourtts}\end{tabular} & \begin{tabular}[c]{@{}l@{}}474 (all) \\ 289 (en)\end{tabular} & 2.8736       &       3.6034 $\pm$ 0.29         &   0.4621 \\ \hline
YourTTS (Exp. 1)   & 245 & 1.091  & 4.102 $\pm$ 0.25   & \textbf{0.7120}\\ \hline
YourTTS (Exp. 2)      & \begin{tabular}[c]{@{}l@{}} 27k (all) \\ 14k (en)\end{tabular} & 3.4803 & 3.6821 $\pm$  0.29  & 0.5651 \\ \hline
XTTS   (Exp. 3)   & \begin{tabular}[c]{@{}l@{}} 27k (all) \\ 14k (en)\end{tabular}   &  \textbf{0.5425}  &  4.007 $\pm$ 0.25 &  0.6423\\ \hline
\end{tabular}
}
\vspace{-0.6cm}
\end{table}

 Table \ref{tab:results-en} presents  CER, UTMOS, and SECS for all our experiments and related works in the English language. YourTTS monolingual (Exp. 1) presents better results in speaker similarity (SECS) it also shows competitive results in CER and UTMOS metrics. 
 However, it achieved the worst CER among the monolingual models. In fact, YourTTS prosody is not great because it sometimes produces unnatural durations. Comparing Monolingual YourTTS (Exp. 1) with the original multilingual YourTTS we can see a huge improvement. In that way, confirming the over-fitting issue, and showing that previous models miss-compared their model with YourTTS. Comparing Monolingual YourTTS (Exp. 1) with the YourTTS trained on the XTTS dataset (Exp. 2) we can see a huge gap, in all the metrics indicating that comparing multilingual models with monolingual models is not fair. It also shows that YourTTS had difficulties to learn all 16 languages well. XTTS model (Exp. 3) achieved the better CER and it achieved competitive results in all the other metrics. It is impressive especially because our model was trained in 16 languages and we are comparing it with related works that were trained only in the English language. Considering the monolingual-related works, HierSpeech++ achieved better results. It achieved better UTMOS, it also achieved the second better SECS and third better CER. Considering the multilingual-related works, Mega-TTS 2 achieved better results than the original YourTTS on English Language.
 
We also measure user preference scores by comparing XTTS with HierSpeech++ and Mega-TTS 2 models. Following \cite{kim2023p}, We evaluate the preference for naturalness, acoustic quality, and human likeness using a comparative mean opinion score (CMOS). Preference tests for speaker similarity are reported using comparative speaker similarity mean opinion score (SMOS). SMOS evaluators are provided with the speaker reference used to generate the model outputs. The CMOS and SMOS values range on a gradual scale varying from -2 (meaning that XTTS is worse than the other model) to +2 (meaning the opposite). We obtain evaluation scores with a minimum of 8 samples from each evaluator with at least 15 evaluators per comparison experiment. Table \ref{tab:user-results-en} demonstrates that XTTS exhibits significantly better results in terms of naturalness, acoustic quality, and human likeness (CMOS) than previous works. It also shows that XTTS is a little worse than previous models in terms of speaker similarity (SMOS). We think that this is expected due to the complexity of massive multilingual training. These results are also aligned with the objective evaluation presented in Table \ref{tab:results-en}.

\begin{table}[]
\caption{User preference scores by comparing XTTS with HierSpeech++ and Mega-TTS 2 models.}
\label{tab:user-results-en}
\centering
\resizebox{0.40\textwidth}{!}{%
\begin{tabular}{c|c|c}
\hline
\textbf{Comparison}    & \textbf{CMOS($\uparrow$)} & \textbf{SMOS($\uparrow$)} \\ \hline
XTTS vs HierSpeech++ &  0.41 $\pm$ 0.26          &  -0.31 $\pm$ 0.36        \\ \hline
XTTS vs Mega-TTS2    & 0.92 $\pm$ 0.22         & -0.39 $\pm$ 0.38       \\ \hline
\end{tabular}
}
\vspace{-0.4cm}
\end{table}


\vspace{-0.08cm}
\subsection{Multilingual evaluation}

For Multilingual evaluation, we compared YourTTS and XTTS  trained on the XTTS dataset (respectively, Exp. 2 and Exp. 3) with the original Mega-TTS 2 model. Table \ref{tab:results-all} presents CER and SECS for XTTS, YourTTS, and Mega-TTS 2 models.  XTTS model was able to achieve better CER and speaker
similarity in almost all languages.

\begin{table}[ht!]
\caption{CER and SECS for YourTTS (Exp. 2), XTTS, and Mega-TTS 2 models for all supported languages.}
\label{tab:results-all}
\centering
\resizebox{0.45\textwidth}{!}{%
\begin{tabular}{l|cc|cc|cc}
\hline
\multirow{2}{*}{Lang.} &
  \multicolumn{2}{c|}{\textbf{YourTTS}} &
  \multicolumn{2}{c|}{\textbf{XTTS}} &
  \multicolumn{2}{c}{\textbf{Mega-TTS 2}} \\ \cline{2-7} 
 &
  \multicolumn{1}{c|}{\textbf{CER($\downarrow$)}} &
  \textbf{SECS($\uparrow$)} &
  \multicolumn{1}{c|}{\textbf{CER($\downarrow$)}} &
  \textbf{SECS($\uparrow$)} &
  \multicolumn{1}{c|}{\textbf{CER($\downarrow$)}} &
  \textbf{SECS($\uparrow$)} \\ \hline
ar    & \multicolumn{1}{c|}{11.1713} & 0.4400  & \multicolumn{1}{c|}{\textbf{3.3503}} & \textbf{0.5007} & \multicolumn{1}{c|}{-} & - \\ \hline
cs    & \multicolumn{1}{c|}{4.0174}  & 0.4496 & \multicolumn{1}{c|}{\textbf{1.3295}} & \textbf{0.4655} & \multicolumn{1}{c|}{-} & - \\ \hline
de    & \multicolumn{1}{c|}{\textbf{2.2411}}  & 0.4612 & \multicolumn{1}{c|}{{3.1694}} & \textbf{0.5175} & \multicolumn{1}{c|}{-} & - \\ \hline
en    & \multicolumn{1}{c|}{2.9727}  & 0.5651 & \multicolumn{1}{c|}{\textbf{0.5425}} & 0.6423 & \multicolumn{1}{c|}{1.4269}  &  \textbf{0.6428} \\ \hline
es    & \multicolumn{1}{c|}{\textbf{1.0926}}  & 0.4879 & \multicolumn{1}{c|}{{1.4606}} & \textbf{0.5371} & \multicolumn{1}{c|}{-} & - \\ \hline
fr    & \multicolumn{1}{c|}{3.3965}  & 0.4376  & \multicolumn{1}{c|}{\textbf{1.4937}} & \textbf{0.4799} & \multicolumn{1}{c|}{-} & - \\ \hline
hu    & \multicolumn{1}{c|}{4.5098}  & \textbf{0.4819} & \multicolumn{1}{c|}{\textbf{1.4622}} & {0.4570} & \multicolumn{1}{c|}{-} & - \\ \hline
it    & \multicolumn{1}{c|}{1.7010}  & 0.4520 & \multicolumn{1}{c|}{\textbf{0.7982}} & \textbf{0.5008} & \multicolumn{1}{c|}{-} & - \\ \hline
ja    & \multicolumn{1}{c|}{10.2808}   & 0.4873 & \multicolumn{1}{c|}{\textbf{5.3748}} & \textbf{0.5207} & \multicolumn{1}{c|}{-} & - \\ \hline
ko    & \multicolumn{1}{c|}{8.8567}  & \textbf{0.4836} & \multicolumn{1}{c|}{\textbf{4.0647}} & {0.4760} & \multicolumn{1}{c|}{-} & - \\ \hline
nl    & \multicolumn{1}{c|}{3.4228}  & 0.4269 & \multicolumn{1}{c|}{\textbf{0.946}}  & \textbf{0.4825} & \multicolumn{1}{c|}{-} & - \\ \hline
pl    & \multicolumn{1}{c|}{1.5925}  & 0.4561 & \multicolumn{1}{c|}{\textbf{0.7593}} & \textbf{0.4833}  & \multicolumn{1}{c|}{-} & - \\ \hline
pt    & \multicolumn{1}{c|}{1.5481}  & 0.4693 & \multicolumn{1}{c|}{\textbf{1.1068}} & \textbf{0.5033}  & \multicolumn{1}{c|}{-} & - \\ \hline
ru    & \multicolumn{1}{c|}{2.8566}  & 0.4606 & \multicolumn{1}{c|}{\textbf{0.932}}  & \textbf{0.5012} & \multicolumn{1}{c|}{-} & - \\ \hline
tr    & \multicolumn{1}{c|}{2.6367}  &  0.4855 & \multicolumn{1}{c|}{\textbf{1.042}}  & \textbf{0.5031} & \multicolumn{1}{c|}{-} & - \\ \hline
zh-cn & \multicolumn{1}{c|}{14.4220} & 0.4825 & \multicolumn{1}{c|}{\textbf{5.2016}} & \textbf{0.5023} & \multicolumn{1}{c|}{{6.1031}}  &  0.4529 \\ \hline
Avg.  & \multicolumn{1}{c|}{4.7949}  & 0.4704  & \multicolumn{1}{c|}{\textbf{2.0646}} & \textbf{0.5046} & \multicolumn{1}{c|}{-} & - \\ \hline
\end{tabular}
}
\vspace{-0.4cm}
\end{table}

 \vspace{-0.12cm}
\section{Speaker Adaptation}

The different recording conditions are a challenge for the generalization of the ZS-TTS models \cite{yourtts}. Speakers who have a voice that differs greatly from those seen in training also become a challenge \cite{tan2021survey}. Nevertheless, to show the potential of the XTTS model for adaptation to new speakers/recording conditions, we selected samples of approximately 10 min of speech from well-known or unique-style voices  (e.g. whispering voices) in different languages. We choose 3 speakers of English, 3 speakers of Portuguese, 1 speaker of Chinese, and 1 speaker of Arabic. We fine-tuned using these speakers and we evaluated the model using the cross-lingual approach used in Section \ref{sec:results}; however, we replaced the DAPS speakers with the chosen speakers. The fine-tuned model improves the SECS from 0.5852 to 0.7166 when cloning these voices in a cross-lingual way. It indicates that the XTTS fine-tuning improved the speaker similarity a lot in cross-lingual speaker transfer settings.  The results are available on the demo page\footnote{https://edresson.github.io/XTTS}. 

\vspace{-0.12cm}
\section{Conclusions and future work} \label{sec:conc}

In this work, we presented XTTS, which achieved SOTA results in Multilingual zero-shot multi-speaker TTS in 16 languages. Furthermore, we showed that XTTS can be fine-tuned with a small portion of speech and achieves impressive results in prosody and style mimicking, being able to mimic a whispering voice style in all 16 languages even though it was trained with only 10 minutes of a whispering English voice. The XTTS model is also faster than VALL-E because our encoder produces tokens at a 21.53 Hz frame rate as compared with 75Hz from the VALL-E model.  In future work, we intend to seek improvements to our VQ-VAE component to be able to generate speech with the VQ-VAE decoder instead of using the current XTTS Decoder component. We also intend to disentangle speaker and prosody information to be able to do cross-speaker prosody transfer. 

\vspace{-0.12cm}
\section{Acknowledgments}

We would like to thank all Coqui TTS\footnote{https://github.com/coqui-ai/TTS} contributors, this work was only possible thanks to the commitment of all. Also, we want to thank HierSpeech++, Tortoise, and StyleTTS 2 authors for making their work open-source and easily accessible to the community. In addition, we want to thank Ziyue Jiang, for kindly generating Mega-TTS 2 model samples used in this paper.
\vspace{-0.12cm}

\bibliographystyle{RefStyle}

\bibliography{references}


\end{document}